\documentclass[runningheads,a4paper]{llncs}

\usepackage{amssymb}
\usepackage{graphicx}

\usepackage{url}
\urldef{\mailsa}\path|yodaiken@fsmlabs.com|
\urldef{\mailsc}\path|@fsmlabs.com|
\newcommand{\keywords}[1]{\par\addvspace\baselineskip
\noindent\keywordname\enspace\ignorespaces#1}
\newcommand{\set}[1]{\{#1\}}
\newcommand{\seq}[1]{\langle#1\rangle }
\newcommand{\ess}{\Lambda}

\newcommand{\Ma}{\mathcal{M}}
\newcommand{\Aprod}{\mathcal{A}}

\newcommand{\st}{\mathit{start}}

\newcommand{\Bfunc}{\left\{\begin{array}{ll}}
\newcommand{\Efunc}{\end{array}\right.}
\newcommand{\Beq}{\begin{eqnarray}}
\newcommand{\Eeq}{\end{eqnarray}}

\newcommand*{\xLongleftarrow}{\ensuremath{\Leftarrow\joinrel\Relbar\joinrel\Relbar\joinrel\Relbar\joinrel\Relbar\joinrel\Relbar\joinrel\Relbar\joinrel\Relbar\joinrel\Relbar}}

\newtheorem{thm}{Theorem}
\newtheorem{dfn}{Definition}[section]
\begin{document}

\mainmatter  

\title{Primitive Recursive Presentations of Transducers and their Products}

\titlerunning{Recursive Automata and Products}

%
%
\author{Victor Yodaiken%
\thanks{ This paper replaces multiple earlier rough drafts.  }
}
\authorrunning{Yodaiken}

\institute{FSMLabs Inc. 
2718 Creeks Edge Parkway\\
Austin Texas, 78733, USA
\mailsa\\
\url{http://www.yodaiken.com}}

%
%

\toctitle{Recursive Transducers}
\tocauthor{Yodaiken}
\maketitle

\begin{abstract}
Methods for specifying Moore type state machines (transducers) abstractly via
primitive recursive string functions are discussed.
The method is mostly
of interest as a concise and convenient way of working with 
the complex state systems found in computer programming and engineering, but  a short section indicates connections to algebraic automata theory and the theorem of Krohn and Rhodes. The techniques are shown to allow concise definition
of system architectures and the compositional construction
 of parallel and concurrent systems.
\end{abstract}

\section{Introduction}

The engineering disciplines of programming and computer system design have
been handicapped by the practical limitations of mathematical techniques for
specifying complex discrete state systems. While finite automata are the natural
basis for such efforts, the traditional state-set presentations of automata
are convenient for only the simplest systems as well as for classes of systems,
but become awkward when state sets are large, when 
behavior is only partially specified, and for compositional systems.
Furthermore, it would be nice to be able to parameterize
automata so that we can treat, for example, an 8bit memory as differing from 
a 64bit memory in only one or a few parameters. These problems can all be
addressed by using a recursive function presentation of automata that is 
introduced here.

General automata have long been understood to be a class of
functions from finite
strings of input symbols to finite strings of output symbols\cite{Arbibabs}
but for specifying computer systems it is more useful to consider functions
from finite strings of inputs to individual outputs. The intuition is that
each string describes a path from the initial state to some ``current" state
and the value of the function is the output of the system in the ``current"
state. If $A$ is an alphabet of input events and $X$ is a set of possible 
outputs, let $A^*$ be the set of finite strings over $A$ including the
empty string $\ess$ and then a function  $f:A^*\to X$ defines a
relationship between input sequences and outputs. 
These functions can be shown to be strongly 
equivalent to (not necessarily finite) Moore type automata\cite{Moore}
while abstracting out details that are not interesting for our purposes here.
If $a$ is an input and
$w$ is a string, $wa$ is the result of appending $a$ to $w$ and by
defining $f(\ess)=x_0$ and $f(wa) = h(a,f(w))$, we can completely specify
the operation of $f$. 

\begin{center}
\fbox{
\begin{minipage}{4in}
Correspondence between a transducer $M$ and a string function $f$.
\[\begin{array}{c}
\mbox{Input:}w\Rightarrow \mbox{Machine:}M\Rightarrow \mbox{Output:}x\\
f(w)=x
\end{array} \]
\end{minipage}
}
\end{center}

It turns out that a type of simultaneous 
recursion can be used to specify automata products that 
model composition and parallel (and concurrent) state change.
Suppose that $f_1, ... f_n$ are previously defined
string functions, $f_i:A_i^*\to X_i$ and we wish to combine these into
a system where inputs from some alphabet $A$ drive the components forward.
At each step an input $a$ to the composite system will be used to generate
an input sequence $z_i$ for each component $f_i$.  The input sequence for
the component is a function of both $a$ and the feedback, the outputs of
the components. The composition builds a new function
$f$ from  $f_1 \dots, f_n$ plus a 
communication map $g$ and an output map $h$. 
Let $f(w) = h(f_i(u_1)\dots, f_n(u_n))$
where the $u_i$ are themselves primitive recursive functions of $f$ and
$w$. I will write $u_i$ when $w$ is clear from context and use functional
form $u_i(w)$ otherwise. We always require that $u_i(\ess)=\ess$ --- so that in the initial 
state of the composite system every component is in its own initial state.
Let $w \circ z$  be the string obtained by concatenating $w$ and $z$.
The communication map is used as follows:$u_i(wa)= u_i(w) \circ g(i,a,f(w))$.  
The idea is that appending $a$ to $w$ causes
the string $g(i,a,f(w))$ to be concatenated to $u_i$. 

\vspace{0.7cm}

\paragraph{Outline.}

In what follows, I'll give two examples of parallel composition and
then make the correspondence between string functions and transducers
precise and prove the correspondence between the simultaneous recursion scheme given above to a "general product" of automata. The concluding section
looks at some implications 
for the study of automata structure and algebraic automata theory. 
Companion technical reports describe practical use. 

The  two ``factors" case is illustrative.
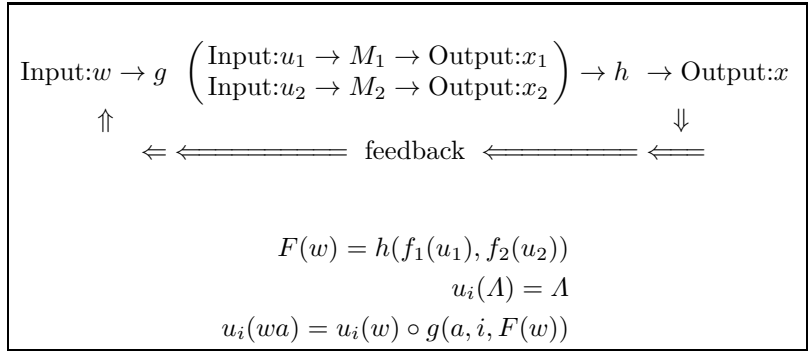
\begin{center}
\fbox{\begin{minipage}{4in}
\[\begin{array}{rcl}\mbox{Input:}w\rightarrow g&\left(\begin{array}{l}
\mbox{Input:}u_1\rightarrow M_1\rightarrow\mbox{Output:}x_1\\
\mbox{Input:}u_2\rightarrow M_2\rightarrow\mbox{Output:}x_2
\end{array}\right)\rightarrow h&\rightarrow\mbox{Output:}x\\
\Uparrow\qquad& &\quad \Downarrow\\
\Leftarrow&\xLongleftarrow\joinrel\Relbar\mbox{ feedback }\xLongleftarrow&\Longleftarrow\joinrel\Relbar\end{array}\]

\Beq F(w) = h( f_1(u_1), f_2(u_2))\nonumber \\
     u_i(\ess) = \ess\nonumber \\
     u_i(wa) = u_i(w)\circ g(a,i,F(w))\nonumber \Eeq
\end{minipage}
}
\end{center}

\paragraph{Example: Stack}
By way of illustration consider a parallel implementation of a stack.
\Beq Stack_n(w) = (S(u_1) \dots , S(u_n))\Eeq
where each $S(za)=a$ so that the  $n$ factors are
 are simple storage cells. Let's 
have a special value so we can spot empty cells $S(\ess)=EMPTY$ and 
have some $a=EMPTY$ in the storage cell alphabet. The alphabet
of the stack is $PUSH[v] : v\in A_{storage}$ and $POP$. Then define the $u_i$
\Beq
u_i(\ess)=\ess\\
u_i(wa)= u_i(w)\circ \Bfunc 
	\seq{v}&\mbox{if }i=1 \mbox{ and }a=PUSH[v]\\
	\seq{EMPTY}&\mbox{if }i=n \mbox{ and }a=POP\\
	\seq{S(u_{i-1}(w))}&\mbox{ if }i>1\mbox{ and }a=PUSH[v]\\
	\seq{S(u_{i+1}(w))}&\mbox{ if }i<n\mbox{ and }a=POP\Efunc
\Eeq
Then define $Top(w) = S(u_1)$ and 
\[Empty(w)=\Bfunc
	1&\mbox{if }S(u_1)=EMPTY\\
	0&\mbox{otherwise}.\Efunc \]
and
\[Full(w)=\Bfunc
	1&\mbox{if }S(u_n)\neq EMPTY\\
	0&\mbox{otherwise}.\Efunc \]

\paragraph{Example: Network}
A computer on a network might, from the outside, appear to have an alphabet
consisting of $RECV[m]$, $TRANSMIT[m]$, for $m$ in a set of possible
messages and $TICK$ to indicate passage of time.  Say $D$ is a networked
computer if $D(w)\in \set{(m,c): m\in Messages\cup \set{NULL}, c\in \set{ ready, busy})}$ where $D(w)=(x,y)$ tells us that $D$ is trying to send message $x$
(or not sending any message if $x=NULL$) and that $D$ is or is not ready to 
accept a message. For simplicity assume a broadcast network and then define
\[  N(w) = (D_1(u_1)\dots , D_n(u_n), R(v))\]
where each $D_i$ is a network node and $R$ is an arbiter we can 
define to pick which, if any, node gets to send a message next. Each $D_i$
may be distinct as long as it satisfies the specifications of output values.
\[R(z) \in\set{1\dots n}\]
The alphabet of $N$ can just consist of the single symbol $TICK$.
Let $u_i(wa) = u_i(w)\circ \seq{RECV[m],TICK}$ if $R(v(w))=j$ and 
$D_j(u_j(w))=(m,c)$ and $D_i(u_i(w))=(k,ready)$. Otherwise, just append
$TICK$ to $u_i$. 

If $D_j$ is itself a product, say $D_j(w)= (OS(r_{os}), APP(r_{app})$ then if $w$ is the string parameter to $N$, we can look inside at the value of
$OS(r_{os}(u_i(w)))$.

\section{Basics\label{sec:basics}}

A Moore machine or transducer is usually given by a 6-tuple
$$M=(A,X, S,\st,\delta,\gamma)$$
where $A$ is the alphabet, $X$ is a set of outputs, $S$ is a set of states, $\st\in S$ is the initial
state, $\delta: S\times A\to S$ is the transition function
and $\gamma:S\to X$is the output function.

Given $M$, use primitive recursion on sequences to 
extend  the transition function $\delta$ to $A^*$ by:
\begin{eqnarray}
\delta^*(s,\ess)=s \mbox{ and } \delta^*(s,wa)=\delta(\delta^*(s,w),a).
\end{eqnarray}

So $\gamma(\delta^*(\st,w))$ is
the output of $M$ in the state reached by following $w$ from $M$'s initial
state.
Call $f_M(w) =\gamma(\delta^*(\st,w))$ the 
\emph{representing function} of $M$. 

If $f_M$ is the representing function of $M$, then 
$f'(w)= g(f(w))$ represents $M'$
obtained by replacing $\gamma$ with $\gamma'(s) = g(\gamma(s))$. The state
set of $M$ and transition map remain unchanged. 

The transformation from string function to transducer is also simple.
Given $f:A^*\to X$ define  $f_w(u) = f(w\circ u)$. Let
 $S_f =\set{f_w: w\in A^*}$. 
Say $f$ is finite if and only if $S_f$ is finite. Define
$\delta_f(f_w,a)=f_{wa}$ and define $\gamma(f_w)= f_w(\ess) = f(w)$. Then
with $\st_f = f_\ess$ we have a Moore machine
\[\Ma(f)=\set{S_f,\st_f,\delta_f,\gamma_f}\]
and, by construction $f$ is the representing function for $\Ma(f)$.

A similar construction can be used
to produce a monoid from a string function as discussed below in section
\ref{sec:monoids}.

Any $M_2$ that has $f$ as a representing function can differ
from $M_1=\Ma(f)$ only in names of states and by including unreachable and/or
duplicative states. That is, there may be some  $w$ so that
$\delta_1^*(\st_1,w) \neq \delta_2^*(\st_2,w)$ but since $f_w= f_w$ it must
be the case that the states are identical in output and in the output of any
states reachable from them. If we are using Moore machines to
represent the behavior of digital systems, these differences are not particularly interesting and we can treat $\Ma(f)$ as \emph{the} Moore machine represented
by $f$.

While finite string
functions are the only ones that can directly model digital computer devices
or processes\footnote{There is confusion on this subject for reasons
I cannot fathom, but processes executing on real computers are not Turing machines because real computers do not have infinite tapes and the possibility of 
removeable tapes doesn't make any difference.}, infinite ones are 
often useful in describing system properties.  For example, we may
want $L(\ess)=0$ and $L(wa)= L(w)+1$ and then seek to 
prove for some $P$ that there is a $t_0$ so that whenever
 $L(w\circ z) \geq  L(w)+t_0$ there is a prefix $v$ of $z$ so that 
$P(w\circ v)=0$. In this case, $L$ is an ideal measuring device, not
necessarily something we could actually build.

\subsection{Products}

Suppose we have a collection of (not necessarily distinct) Moore machines
$M_i=(A_i,X_i,S_i,\st_i,\delta_i,\lambda_i)$ for $(0 < i \leq n)$
that are to be connected to construct a new machine with alphabet $A$ 
using a connection map $g$.
The intuition is that when an input $a$ is applied to the system,
the connection map computes a string of inputs for
$M_i$ from the input $a$ and the outputs of the factors (\emph{feedback}).
The general product here is described by Gécseg \cite{Gecseg}.
I have made the connection
maps generate strings instead of single events
so that the factors can run at non-uniform rates.
If $g(i,a,\vec{x})=\ess$, then $M_i$ skips a turn.

\begin{dfn}{\rm
\textbf{General product of automata}\\
Given $M_i=(A_i,X_i,S_i,\st_i,\delta_i,\gamma_i)$ and $h$ and 
$g$ define the Moore machine:
$M = \Aprod_{i=1}^n [ M_i, g,h] = (A,X,S,\st, \delta,\gamma)$\\
\begin{itemize}
\item $S = \set{(s_1 \ldots,s_n): s_i\in X_i}$ and $\st = (\st_1 \ldots ,\st_n)$
\item $X=\set{h(x_1 \ldots  ,x_n): x_i\in X_i}$ and $\gamma((s_1 \ldots , s_n))= h(\gamma_1(s_1)\ldots ,\gamma_n(s_n) )$.
\item $\delta((s_1 \ldots , s_n),a) = (\delta^*_1(s_1,g(1,a,\gamma(s))) \ldots , \delta^*_n(s_n,g(n,a,\gamma(s))))$.
\end{itemize} }
\end{dfn}

One thing to note is that the general product, in fact any product of automata, is likely to produce a state set that contains unreachable states. The
string function created by simultaneous recursion represents the minimized
state machine as well. The possible ``blow up" of unreachable and duplicate states is not a problem for composite recursion although it vastly complicates
work with state-set representations.

\begin{thm}\label{thm:main}{\rm
\textbf{If} each $f_i$ represents $M_i$
and $f(w) = h(f_1(u_1)\dots,f_n(u_n))$\\
and $u_i(\ess)=\ess$ \\
and $u_i(wa) =u_i(w)\circ g(i,a,f(w))$\\
and $M= \Aprod_{i=1}^n [M_i, h,g]$
\textbf{then} $f$ represents $M$
}\end{thm}

\textbf{Proof:} 
Each $f_i$ represents $M_i$ so
\begin{eqnarray}f_i(z) = \gamma_i(\delta^*_i(\st_i,z))\label{eqn:c}\end{eqnarray}
But $\gamma(\delta^*(\st,w)) = h(\gamma(s)) = h(\dots \gamma_i(\delta_i^*(\st_i,w_i))\dots )$
for some $w_i$.  All we have to show is that 
\Beq \delta^*(\st,w)= (\dots \delta_i^*(\st_i,u_i(w))\dots ) \label{eqn:a}\Eeq
and then we have 
\[\gamma(\delta^*(\st,w)) = h(\dots \gamma_i(\delta_i^*(\st_i,u_i(w)))\dots ). \]
It follows immediately that 
\[\gamma(\delta^*(\st,w)) = h(\dots f_i(u_i(w)))\dots ) = f(w)\]
Equation \ref{eqn:a} can be proved by induction on $w$. 
Since $u_i(\ess)=\ess$ the base case is obvious. Now suppose that equation
$\ref{eqn:a}$ is correct for $w$ and consider $wa$.

Let $\delta(\st,w)=s=(s_1 \ldots , s_n)$ and let $u_i(w)=z_i$. Then, by
the induction hypothesis $s_i = \delta_i^*(\st_i,z_i)$, and, by the argument
above $\gamma(\delta^*(\st,w)) = f(w)$. So:\\
\begin{eqnarray}
\delta^*(\st,wa) = \delta(\delta^*(\st,w),a)\\
= \delta(s,a)\\
= (\dots \delta_i^*(s_i,g(i,a,\gamma(s)))\dots)\\
= (\dots \delta_i^*(\delta_i^*(\st,u_i(w)),g(i,a,f(w)))\dots)\\
= (\dots \delta_i^*(\st,u_i(w)\circ g(i,a,f(w)))\dots)\\
= (\dots \delta_i^*(\st,u_i(wa))\dots)\end{eqnarray}
proving \ref{eqn:a} for $wa$.

It follows directly that if $M$ is represented by $f$, and $f$ is defined
by simultaneous recursion, then $f$ can also
be defined by single recursion --- although such a definition may be 
impractical because of the large state set size.

\section{More on Representation and Some Algebra\label{sec:algebra}} 

A number of results follow from theorem \ref{thm:main}.
\begin{thm}
For $M$ and $f$ constructed as products as above in theorem \ref{thm:main}.
\begin{itemize}
\item There are an infinite number of distinct products $M'= \Aprod_{i=1}^k [N_i,g_i]$ so that $f$ represents $M'$ as well as $M$.
\item If all of the $M_i$ are finite state, $M$ is finite state (by construction).
\item If all of the $f_i$ are finite state, $f$ is finite state (
since it represents a finite state Moore machine).
\item If $f$ is finite state then there is some $M' = \Aprod_{i=1}^k r[Z_i,g,h]$
where $f$ represents $M'$ and each $Z_i$ is a 2 state Moore machine. In fact
$k = \lceil\log_2(|S_{M'}|)\rceil$. This is simple binary encoding.
\end{itemize}
\end{thm}

%

\subsection{Monoids\label{sec:monoids}}
If $f:A^*\to X$ then say $w \equiv_f u$ iff $f(z\circ w\circ y) = f(z\circ u\circ y)$ for all
$z,y\in A^*$.
Let $[w]_{/f}=\set{ u\in A^*, u \equiv w}$. Then define
$[w]_{/f}\cdot [z]_{/f}= [w\circ z]_{/f}$. The set of these classes with
$\cdot$
comprises a monoid where
$[w]_{/f}\cdot [\ess]_{/f} = [w]_{/f}$ for the required 
identity. Say that this monoid is the monoid determined by $f$. Recall
the construction of states from string functions above and the set $S_f$
consisting of all the functions $f_w$ so that $f_w(z)= f(w\circ z)$. Note
that if $v,z\in [w]_{/f}$ it must be the case that for any string $r$
$f_{r\circ z} = f_{r\circ v}$. So it is possible to associate each 
$[w]_{/f}$ with a map from $S_f\to S_f$ where $f_r\mapsto  f_{r\circ z}$
for any $z$ in $[w]_{/f}$. As a result, whenever $S_f$ is finite, there
are only a finite number of maps $S_f\to S_f$ so the monoid determined by
$f$ must also be finite.

Suppose $f(w)= h(f_1(u_1) \ldots , f_n(u_n))$ so that
$u_i(wa)= u_i(w)\circ z_i$ where $z_i$ only depends on the feedback
from factors indexed by $j< i$. That is, there are 
$r_1 \ldots  r_n$ so that $z_1= r_1(a)$ and 
 $z_{i+1} = r_{i+1}(a,f(w,1) \ldots ,f(w,i))$.
In this case $f$ is 
constructed in cascade where information flows only in one direction
and the results of Krohn-Rhodes theory\cite{Holcombe,Ginzburg} will apply.

If $f$ is finite
and represents a state machine with $k$ states and each of the $f_i$ are
finite with $k_i$ states in the represented state machine, then if
$\Sigma_{j=1}^{j\leq n}k_j < k$ the factorization is an implementation 
of $f$ by essentially simpler string functions --- and it corresponds to 
a factorization of the monoid of $f$ into simpler monoids.

Let $T_n(\ess)=0$ and $T_n(wa) = T(w)+1\bmod n$. Now define $G_n$ as a
cascade of $T_2$'s as follows:
\begin{eqnarray}
G_n(w) = (T_2(u_1)\dots ,T_2(u_n))\\
u_1(wa) = u_1(w)\circ \seq{a} = wa\\
u_{i+1}(wa) = u_{i+1}(w)\circ \Bfunc
\ess&\mbox{if } \exists j< i, T_2(u_j(w))=0\\
\seq{a}&\mbox{otherwise}\Efunc \end{eqnarray}
This is called a ``ripple carry adder" in digital circuit engineering: each counter increments only if the ``carry" is propagating through all lower order counters. Put $H_n(w) = \Sigma_{i=1}^{i\leq n}T_2(u_i)\times 2^{i-1}$ where the 
$u_i$ are as defined for $G_n$. Then $H_n = T_{2^n}$ and you cannot make
a $G_n$ which counts mod any number other then $2^n$. 
Otherwise, the underlying monoid of
$T_k$ has a simple group factor (a prime cyclic group)
and those cannot be factored into smaller elements without some feedback.

While the cascade decompositions may simplify the interconnect in one way, they
do not necessarily indicate the most efficient or interesting
decomposition in practice. Cascades are good designs for "pipelined"
execution but may be slow if we have to wait for the data to propagate to
the terminal element. And group qualities in data structures can correspond to
"undo" properties. For example, consider a circular buffer - 
like those commonly used for UNIX type fifos/pipes. The idea
is that "write" operations push data into the pipe and "read" operations remove
data in order of the "writes". The memory used to hold the data is allocated
in a cycle. One way to implement such a buffer is to decompose it into an 
array of $k$ memory locations and a mod $k$ counter. A write operation causes
an increment of the counter and a store of data in the appropriate memory location. The increment has an inverse, the write does not. But the result is that
a write can be ``forgotten". Perhaps factoring off group-like components will
reveal other possibilities for this type of partial inverse. 

\bibliography{all}
\bibliographystyle{plain}
\end{document}